# An electrokinetic route to giant augmentation in load bearing capacity of compliant microfluidic channels

Siddhartha Mukherjee,[1], Jayabrata Dhar[2], Sunando Dasgupta[1,3], Suman Chakraborty[1,4]

[1]*Advanced Technology Development Center, Indian Institute of Technology Kharagpur, Kharagpur, India-721302*

[2]*Université de Rennes 1, CNRS, Géosciences Rennes UMR6118, Rennes, France*

[1,3]*Department of Chemical Engineering, Indian Institute of Technology Kharagpur, Kharagpur, India-721302*

[1,4]*Department of Mechanical Engineering, Indian Institute of Technology Kharagpur, Kharagpur, India-721302*

## Abstract

The performances of lubricated systems widely used in natural, biological, and artificial settings are traditionally dictated by their load bearing capacities. Here we unveil that, by exploiting a unique coupling between interfacial electro-mechanics, hydrodynamics and substrate compliance, it is plausible to realize a massive augmentation in the load bearing capacities of compliant microfluidic channels. Our analysis demonstrates that the interplay between wettability and charge modulation in association with the solution chemistry and surface compliance results in this remarkable phenomenon. These results are likely to open up novel design paradigms of augmenting the load bearing capacities of miniaturized bio-mimetic units through the realization of a symmetry breaking phenomenon triggered by asymmetries in electro-mechanical and hydrodynamic transport over interfacial scales.

* E-mail address for correspondence: suman@mech.iitkgp.ernet.in

# 1. Introduction

Lubricated systems are pretty common in the mechanical as well as in the physiological world (Abkarian *et al.* 2002; Beaucourt *et al.* 2004; Christov *et al.* 2018; Elbaz & Gat 2014, 2016; Grodzinsky *et al.* 1978; Martin *et al.* 2002; Ozsun *et al.* 2013; Salez & Mahadevan 2015; Szeri *et al.* 2008; Tanner 1966; Urzay 2010; Urzay *et al.* 2007). These systems find extensive applications in micro-fabrication purposes, spatio-temporal flow control in narrow confinements and bears broader applicability when coupled with interfacial phenomenon like electrokinetics (Dendukuri *et al.* 2007; Hardy *et al.* 2009; Kartalov *et al.* 2007; Mukherjee *et al.* 2013; Panda *et al.* 2009; Seker *et al.* 2009). Lubricated systems are often attributed by a flexible nature of the solid surfaces interfacing with the fluid media, mimicking physiologically relevant bio-mimetic processes to a large extent (Chakraborty S. 2010; Matse *et al.* 2018; McDonald & Whitesides 2002; Nase *et al.* 2008; Raj & Sen 2016; Raj *et al.* 2017). Effectiveness of such a system is dictated by its load bearing capacity which quantifies the maximum load that it can withstand (Chakraborty & Chakraborty 2011; Jahn & Klein 2018; Jones *et al.* 2008; Naik *et al.* 2017; Selway & Stokes 2014; Skotheim & Mahadevan 2004, 2005; Style *et al.* 2013; Wakelin 1974; Wang *et al.* 2015).

Flexible lubricated systems also feature substrate wettability and surface charge as two typical attributes especially in nano- or micro-scale transport with biological interfaces. (Ajdari 1995, 1996, 2000; Bahga *et al.* 2010; Chang & Yang 2008; Chen & Cho 2007; Glasgow *et al.* 2004; Mandal *et al.* 2015; Squires & Bazant 2004; Stone *et al.* 2004; Stroock *et al.* 2000, 2002) Despite this obvious confluence, charged fluidic interfaces, typically formed due to electrical double layer (EDL) phenomena, (Ajdari 1995, 1996, Bandopadhyay & Chakraborty 2011, 2012; Bazant *et al.* 2004; Boyko *et al.* 2016; Chakraborty 2006; Das *et al.* 2006; Das & Chakraborty 2010; Ghosal 2002; Ghosh *et al.* 2016; Hunter 1981; Mukherjee *et al.* n.d., 2017b; R. F. Probstein 1994; Zholkovskij *et al.* 2003) have not yet been harnessed in an effort to augment the load bearing capacities of compliant microfluidic units of designed wettability variations. Here, by exploiting a unique interplay of patterned surface charge, patterned surface wettability, and substrate compliance, we bring out exclusive possibilities of obtaining massive augmentations in the resultant load bearing capabilities of flexible microfluidic units. We attribute these findings to a symmetry breaking phenomenon brought about by the asymmetric boundary conditions in the interfacial slip and zeta potential, leading to a resultant additional lift force. These results may turn out to be of immense consequence in exploring the possibilities of giant augmentations in load bearing capacities of biologically relevant miniaturized systems that often possess inhomogeneities in surface charges and surface wettabilities, (Bhattacharya *et al.* 2005; Bonaccurso *et al.* 2003; Dong *et al.* 2010; Ghosh & Chakraborty 2012, 2013; Guo *et al.* 2005; Jin *et al.* 2005; Kogoma *et al.* 1987; Love *et al.* 2005; Sun *et al.* 2004; Wu & Shi 2005; Zheng *et al.* 2006) either as a natural artifact of obvious imperfections, or as an engineered approach towards bringing in asymmetry in the resultant transport characteristics, giving rise to novel fluid-structure interactions that have hitherto remained unexplored.

## 2. Problem Formulation

Figure 1 depicts the schematic of the present problem and represents the electroosmotic flow of a binary symmetric electrolyte (1:1) through a deformable parallel-plate microchannel having length $L$ and half-channel height $H$ where $H \ll L$. Here we have chosen a rectangular Cartesian co-ordinate system where *x* and *y* are the longitudinal and transverse co-ordinates with the channel centreline at the entrance being the origin. Both the channel ends are maintained at isothermal and isobaric conditions. In this analysis, the channel walls are considered to be compliant in nature thus involving solid-fluid interaction in narrow confinements. Also, the channel walls are subjected to patterned wettability gradients which we have implemented here by imposing asymmetric modulation in the slip length as $l_s \left[ 1 + \delta \cos(q x + \theta) \right]$ on the top plate and $l_s \left[ 1 + n \delta \cos(q x + \theta) \right]$ (Ghosh & Chakraborty 2012, 2013) on the bottom where $l_s$ is the slip length, $\delta$ modulation of the axially varying component, *n* asymmetry factor, *q* patterning frequency of modulation and $\theta$ is the phase difference (or phase shift) angle. In this context, it is interesting to mention that some previous research works (Baudry *et al.* 2001; Choi *et al.* 2003; Galea & Attard 2004; Ng & Chu 2011; Ng & Wang 2011; Tretheway & Meinhart 2010) have demonstrated that depending on the degree of roughness, hydrodynamic slip length may vary from few nanometres to few microns and remarkably, this length scale may become equivalent to the channel dimensions typically employed in micro and nanofluidic applications. The patterning in the slip length creates a designed variation in the wall-adjacent velocity gradient thereby imposing relatively hydrophobic behavior in some axial locations and relatively hydrophilic behavior in other locations. A electrical potential field $\phi$ is applied along the axial direction resulting in an electric field as $\mathbf{E} = -\nabla\phi$. The surface potential at the fluid-surface interface, also known as the zeta potential, is chosen as $\zeta\left[1+\gamma\cos(qx+\alpha)\right]$ and consists of both axially varying and invariant components. Here, $\gamma$ is the modulation of the axially varying component of zeta potential. The combined action of the wettability and potential modulation induces pressure gradient in the longitudinal direction and the compliant nature of the channel walls respond elastically to it; thereby creating the intended intricate coupling between electrokinetic, surface and material properties in narrow confinements. The substrate undergoes deformation because of this implicitly induced pressure imbalance, and the elastic response, therefore, becomes a strong function of this modulations. The degree of deformation depends strongly on the involved flow physics and the corresponding modulus of elasticity of the substrate. More the degree of hydrodynamic and electrokinetic perturbation, more is the difficulty to maintain the original separation between the two plates thus leading to higher generation of lift force and the associated load bearing capacity of the device.

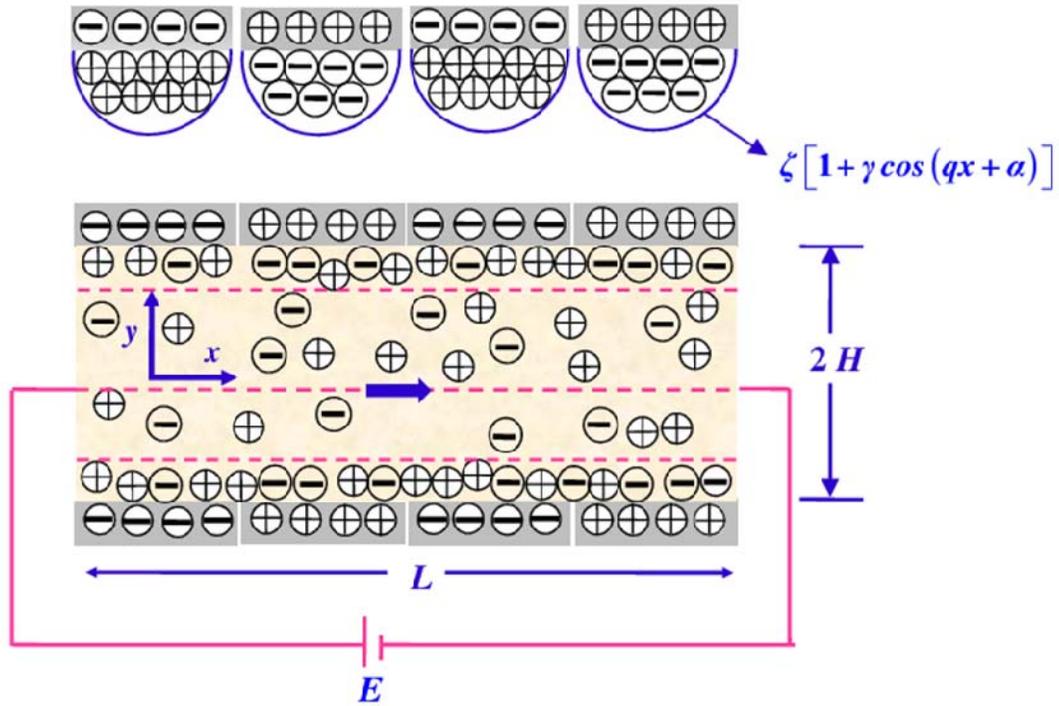

FIGURE 1. Schematic depiction of the physical problem.

The flow in the channel follows low Reynolds number hydrodynamics and is considered to be steady and laminar. Using aforementioned assumptions, the continuity and momentum equations are written as follows

$$\left.\begin{array}{l} \nabla \cdot \boldsymbol{v} = 0 \\ 0 = -\nabla p + \nabla \cdot \boldsymbol{\tau} + \boldsymbol{F}_b \end{array}\right\} \quad (1)$$

where $p$ is the hydrodynamic pressure, $\boldsymbol{v}$ velocity vector, $\boldsymbol{\tau}$ stress tensor, and $\boldsymbol{F}_b$ is the electrokinetic body force given by $\boldsymbol{F}_b = \rho_e \boldsymbol{E}$ where $\rho_e$ is the excess charge density and $\boldsymbol{E}$ is the electric field given by $\boldsymbol{E} = -\nabla \phi$ with $\phi$ being the applied electric potential. Here, the flow is actuated by means of axially applied electric potential, where an electrical double layer (EDL) is formed adjacent to the substrate. Basically, the chemical state of surface undergoes significant alteration upon contact with electrolyte solution which is accompanied by the release of counter-ions for maintaining the electro-neutrality of the system. As a result, the fluid layer in the vicinity of the charged surface possesses a local charge distribution despite being electrically neutral in overall. This charged layer adjacent to the charged surface is termed as EDL. (Chakraborty & Srivastava 2007; Das & Chakraborty 2011; Talapatra & Chakraborty 2008)

The potential distribution in the EDL is governed by the Poisson equation $\nabla^2 \phi = -\rho_e / \varepsilon$ where $\varepsilon$ is the permittivity of the medium. This distribution is also considered to follow the Boltzmann distribution along with the Debye-Hückel linearization approximation (which holds well for low values of surface potential $(\zeta)$, i.e. $\zeta < 25$ mV). (Ajdari 1995, 1996) The charge distribution essentially becomes independent of the electric field with the consideration of weak electric field application and is governed by wall potential (i.e. zeta potential). Therefore, one may express $\phi$ as the linear superposition of two functions, i.e., $\phi(x,y) = \phi_{ext}(x) + \psi(x,y)$ where $\phi_{ext}(x)$ is externally applied electric potential and $\psi(x,y)$ being the charge distribution within the EDL. For, $H \ll L$, $\partial \psi / \partial x \ll \partial \psi / \partial y$ and $\psi(x,y)$ can be approximated as $\psi(y)$ only. Also, $d^2\phi_{ext}/dx^2$ becomes negligible as compared to $d^2\psi/dy^2$ for $H \ll L$ and hence, the Poisson equation gets simplified to the form $d^2\psi/dy^2 = -\rho_e/\varepsilon$.

As far as the deformation of the solid substrate is concerned, one needs to employ the stress balance equation given by (Chakraborty & Chakraborty 2011; Skotheim & Mahadevan 2005; Steinberger *et al.* 2008)

$$\nabla \cdot \sigma = 0 \qquad (2)$$

where $\sigma$ is the stress tensor for solid given by $\sigma = G(\nabla \vec{J} + \nabla \vec{J}^T) + \lambda \nabla \cdot \vec{J}[I]$ with $\vec{J} = [J_x, J_y]^T$ representing the displacement field vector. Here, $G$ and $\lambda$ are Lamé constants for an isotropic, linearly elastic solid. equation (2) subjected to lubrication approximation assumption results

$$\frac{\partial^2 J_y}{\partial y^2} = 0 \qquad (3)$$

Equation (3) is subjected to the stress balance at the compliant layer-fluid interface (i.e. at $y = H - H_c$, $\sigma \cdot \hat{n}\hat{n} = -p$) and no displacement at the compliant layer-solid interface (i.e. at $y = H$, $J_y = 0$). The resulting displacement field is given by (Chakraborty & Chakraborty 2011; Naik *et al.* 2017)

$$J_y = -\left(\frac{H_c}{2G + \lambda}\right) p \qquad (4)$$

where $H_c$ is the thickness of compliant layer and $\left(\frac{H_c}{2G + \lambda}\right) = S^{-1}$; where $S$ represents the stiffness of the microchannel. Now, the governing equations are subjected to the modulated slip and zeta potential boundary conditions as

$$\left.u\right|_{y=\pm H} = \mp l_s \left[1 + n\delta \cos(qx + \theta)\right] \left.\frac{\partial u}{\partial y}\right|_{y=\pm H} \quad \text{where} \quad \begin{cases} n = 0, & \text{at } y = H \\ n = 1, & \text{at } y = -H \end{cases}$$
$$\text{and} \quad \left.\psi\right|_{y=\pm H} = \zeta \left[1 + \gamma \cos(qx + \alpha)\right] \qquad (5)$$

The analytical approach for the solution procedure along with the solution are discussed in detail in the following section.

## 3. Analytical Solution Procedure

To appreciate the effect of both wettability $(\delta)$ and potential modulations $(\gamma)$ on the resulting flow and potential field, we have employed an asymptotic approach using regular perturbation technique where any variable $\chi$ can be expressed in the following way

$$\chi = \chi_0 + \delta(\chi_{11}) + \gamma(\chi_{12}) + \delta^2(\chi_{21}) + \gamma^2(\chi_{22}) + \delta\gamma(\chi_{23}) + \ldots \qquad (6)$$

We have performed this asymptotic analysis correct up to second order. Considering terms beyond second order will not have any difference qualitatively but unnecessarily complicates the theoretical framework, and hence, these terms are neglected for mathematical simplification. Now, the momentum equation in presence of electroosmotic body force is rewritten as

$$0 = -\frac{dp}{dx} + \mu \frac{d^2 u}{dy^2} + \rho_e E \qquad (7)$$

where $\mu$ is the viscosity of fluid and $\rho_e = -\varepsilon(d^2\psi/dy^2)$. Upon Debye-Hückel approximation, $d^2\psi/dy^2$ can be written as $\kappa^2\psi$ where $\kappa$ is the reciprocal of Debye length defined by $\kappa = \sqrt{\dfrac{2n_0 z^2 e^2}{k_B T}}$ where $n_0$ is electrolyte concentration, $z$ valence of ions, $e$ electronic charge, $k_B$ Boltzmann constant and $T$ absolute temperature. Using the expansion of equation (6), equation (7) is rewritten as

$$\begin{aligned}0 = &-\frac{d}{dx}\left(p_0 + \delta p_{11} + \gamma p_{12} + \delta^2 p_{21} + \gamma^2 p_{22} + \delta\gamma p_{23}\right) \\ &+ \mu \frac{d^2}{dy^2}\left(u_0 + \delta u_{11} + \gamma u_{12} + \delta^2 u_{21} + \gamma^2 u_{22} + \delta\gamma u_{23}\right) - \kappa^2\left(\psi_0 + \gamma\psi_1\right)\end{aligned} \qquad (8)$$

Here, the potential distribution is only affected due to surface charge modulation (only $\mathrm{O}(\gamma^0)$ and $\mathrm{O}(\gamma^1)$ terms are present). Here, $\mathrm{O}(\delta\gamma)$ term represents the coupling between the slip and potential modulation parameters. Similarly, the boundary conditions described by equation (5) are expanded as

$$\left(u_0 + \delta u_{11} + \gamma u_{12} + \delta^2 u_{21} + \gamma^2 u_{22} + \delta\gamma u_{23}\right)\Big|_{y=\pm H}$$

$$= \mp l_s\left[1 + n\delta\cos(qx+\theta)\right]\frac{\partial}{\partial y}\left(u_0 + \delta u_{11} + \gamma u_{12} + \delta^2 u_{21} + \gamma^2 u_{22} + \delta\gamma u_{23}\right)\Big|_{y=\pm H}, \text{ where } \begin{cases} n=0, & \text{at } y=H \\ n=1, & \text{at } y=-H \end{cases} \quad (9)$$

$$\text{and } \left(\psi_0 + \gamma\psi_1\right)\Big|_{y=\pm H} = \zeta\left[1 + \gamma\cos(qx+\alpha)\right]$$

Here, we have presented the solution for the case when the modulated slip length at the walls are symmetric in nature, i.e. $n=1$. The corresponding expression for asymmetric slip length are too long because of mathematical complexity associated. Hence, these expressions are accessible in the supplementary material. The governing equations along with the solutions for $n=1$ are shown below: For $O(\delta^0, \gamma^0)$, i.e. leading order:

$$\left.\begin{array}{l} \dfrac{\partial u_0}{\partial x} + \dfrac{\partial v_0}{\partial y} = 0 \\[2mm] \mu\dfrac{\partial^2 u_0}{\partial y^2} = \dfrac{dp_0}{dx} + \varepsilon\kappa^2\psi_0 E \\[2mm] \nabla^2\psi_0 = \kappa^2\psi_0 \end{array}\right\} \quad (10)$$

with the boundary conditions $u_0\big|_{y=\pm H} = \mp l_s\left[1 + \delta\cos(qx+\theta)\right]\dfrac{\partial u_0}{\partial y}\bigg|_{y=\pm H}$ and $\psi_0\big|_{y=\pm H} = \zeta$.

The leading order solution represents the condition of perturbation free flow field which implies zero induced pressure gradient. Hence, the effect of surface compliance will not have any role to govern the flow physics in the leading order. Now, the solutions for $O(\gamma^0)$ are given by

$$\psi_0 = \zeta\frac{\cosh(\kappa y)}{\cosh(\kappa H)} \quad \text{and} \quad u_0 = u_{hs}\left\{1 - \frac{\cosh(\kappa y)}{\cosh(\kappa H)}\left[1 + l_s\kappa\tanh(\kappa H)\right]\right\} \quad (11)$$

where $u_0$ represents EOF of an electrolyte in a slit microchannel subjected to wall slip boundary condition and $u_{hs}$ is the characteristic electroosmotic velocity scale defined by $u_{hs} = -\varepsilon E\zeta/\mu$ (also known as Helmholtz-Smoluchowski velocity scale). Similarly, for $O(\delta)$:

$$\psi_{11} = 0; \; u_{11} = \frac{dp_{11}}{dx}\left(\frac{y^2 - H^2 - 2l_s H}{2\mu}\right) + l_s\kappa u_{hs}\tanh(\kappa H)\cos(qx+\theta) \quad (12)$$

This velocity profile can now be integrated across the channel domain to obtain the volume flow rate for $O(\delta)$ as

$$Q_{11} = \int_{-H}^{H} u_{11} \, dy = \frac{dp_{11}}{dx}\left(-\frac{2H^3}{3\mu} - \frac{2l_s H^2}{\mu}\right) + 2l_s \kappa u_{hs} H \tanh(\kappa H) \cos(qx+\theta) \quad (13)$$

One can express the flow physics of the present analysis in terms of the Reynolds equation in association with the lubrication approximation which yields

$$\int_{-H}^{H} \frac{\partial u_{11}}{\partial x} dy + \int_{-H}^{H} \frac{\partial v_{11}}{\partial y} dy = 0 \quad (14)$$

Here $v$-component satisfies the no-penetration condition at the walls (i.e. $v|_{y=\pm H} = 0$). Now, Leibnitz's rule can be applied to equation (14) which results

$$\int_{-H}^{H} \frac{\partial u_{11}}{\partial x} dy = \frac{\partial}{\partial x}\int_{-H}^{H} u_{11} dy - u_{11}(y=H)\frac{\partial H}{\partial x} + u_{11}(y=-H)\frac{\partial(-H)}{\partial x} = 0 \quad (15)$$

The key aspect to note here that the channel height $H(x)$ in equation (15) is kept as a function of the axial coordinate which bring out the effect of channel elasticity unlike the leading order solution. Hence, $H(x)$ can be written as the combination of two parts, $H(x) = h + d(x)$ where $h$ is the invariant part and $d(x)$ is the variation attributed to the channel compliance. Because of the occurrence of the induced pressure gradient, the compliant channel responds elastically, thus creating deformation $(d)$ of the substrate. For small deformation, it can be considered as proportional to the fluid pressure. To obtain $d$, one needs to employ the aforesaid stress balance equation.

Hence, one can rewrite equation (15) considering aforesaid effect in the following way

$$\frac{d^2 p_{11}}{dx^2}\left(-\frac{2h^3}{3\mu} - \frac{2l_s h^2}{\mu}\right) + \left(h + S^{-1} p_{11}\right)\left\{-2l_s u_{hs} q \sin(qx+\theta)\kappa \tanh(\kappa h) + 2\left(\frac{dp_{11}}{dx}\right)^2 \frac{l_s S^{-1}}{\mu}\right\} = 0 \quad (16)$$

where some simplifications are taken into account like $H^3 \approx h^3$, $H^2 \approx h^2$ and $\tanh(\kappa H) \approx \tanh(\kappa h)$. Now, we non-dimensionalise the pertinent variables as

$$\left.\begin{array}{l}\bar{x} = x/L,\ \bar{y} = \bar{h} = h/L,\ \bar{p} = (p - p_{atm})L/\mu u_{hs},\ \bar{\kappa} = \kappa L, \\ \bar{q} = qL,\ \bar{l}_s = l_s/L,\ \bar{\psi} = \psi/\zeta,\ \beta = \mu u_{hs}/S L^2\end{array}\right\}$$

and subsequently equation (17) is expressed in its dimensionless form as

$$a_1 \frac{d^2 \bar{p}_{11}}{d\bar{x}^2} + (\bar{h} + \beta \bar{p}_{11}) \left\{ a_2 \sin(\bar{q}\bar{x} + \theta) + a_3 \left( \frac{d\bar{p}_{11}}{d\bar{x}} \right)^2 \right\} = 0 \qquad (17)$$

where $a_1 = \left( -\frac{2}{3} \bar{h}^3 - 2 \bar{l}_s \bar{h}^2 \right)$, $a_2 = -2 \bar{l}_s \bar{q} \bar{\kappa} \tanh(\bar{\kappa}\bar{h})$ and $a_3 = 2 \bar{l}_s \beta$.

Equation (17) involves a dimensionless parameter $\beta$ to incorporate the effect of surface compliance where $\beta$ is defined as $\beta = \mu u_c / (S L^2)$, which characterizes the relative strength of viscous and elastic forces. Here, $\mu$ is the viscosity of fluid, $u_c$ characteristic velocity scale, $L$ length scale and $S$ is the stiffness of the substrate per unit length, which may vary between $10^4$ to $10^{10}$ N/m$^3$. For typical electroosmotic flows (EOF), $u_c$ is usually chosen as $u_{HS}$ (Helmholtz-Smoluchowski velocity scale), which is of the order of $\sim O(10^{-3})$ ms$^{-1}$, while $\mu$ is usually $\sim O(10^{-3})$ Pa. s. with $L$ being $\sim O(10^{-3})$ m. Therefore, depending on the stiffness of the microchannel $(S)$, $\beta$ may vary between $10^{-4}$ to $10^{-10}$. (Chakraborty & Chakraborty 2011) Since, this formulation involves a number of variables, a list of symbols along with their physical meanings are presented in **Section S3** of the Supplementary Material for clarification.

It is clear that equation (17) is intrinsically non-linear in nature and analytically intractable. Hence, one can evaluate the pressure distribution along the axial direction numerically by employing the well known shooting technique where the channel ends are maintained at atmospheric pressure (i.e. at $x = 0, L; p = p_{atm}$ which implies that at $\bar{x} = 0, 1; \bar{p} = 0$). Once the pressure distribution is known, one can obtain the load bearing capacity $(\tilde{w})$ of the plate per unit width as

$$\tilde{w}_{11} = \frac{w_{11}}{\mu u_{hs}} = \delta \int_0^1 \bar{p}_{11} \, d\bar{x} \qquad (18)$$

Physically, the load bearing capacity of the channel is the quantification of the maximum amount of load that the device can withstand. Any perturbation in the system induces an imbalance in the axial pressure distribution which in turn creates a lift force at the two plates thus making the separation between the plates axially variant. More the degree of perturbation, more is the lift force thus leading to significant enhancement in the load carrying capacity; a fact which is also intuitive from the mathematical description of the load bearing capacity $(\tilde{w})$ in equation (18).

Similar approach is followed to obtain $O(\delta)$ and higher order solutions. Since, the corresponding expressions are too long, these are presented in the **Appendix** section for the sake of conciseness.

## 4. Results and Discussions

Here we attempt a focused representation of the results which are not only experimentally realizable but may also become beneficial in the designing perspective of lubricated systems. So, our main objective is to showcase the variation of the load bearing capacity of the compliant microchannel under different physical situations which are inevitable at some point, not just for the sake of making the analytical framework more challenging such that one can properly design such devices of improved performance.

Towards this, first we have taken into account the ionization process of silica and glass surfaces which takes place when it comes into contact with an electrolyte solution. (Behrens & Grier 2001; Mukherjee *et al.* 2017a; van der Heyden *et al.* 2005) Considering this ionization mechanism and accompanying alteration in the surface potential (This part is demonstrated in **Section S2** of the Supplementary Material), the variation of the load carrying capacity ratio $\left(\tilde{W}/\tilde{W}_{\text{ref}}\right)$ with varying ionic concentration $(n_0)$ is depicted in figure 2. (a) where $\tilde{W}_{\text{ref}}$ is the reference load capacity for a reference $n_0$, i.e., $n_{0,\text{ref}} = 1\,\text{mM}$. For a fixed $n_0$ and $p$H, introducing the slip length modulation factor $\delta$ significantly alters the velocity gradient in the vicinity of the wall. This effect is quantified by Navier's slip condition which shows strong dependence on the surface velocity gradient, and therefore, increasing $\delta$ significantly amplifies the velocity gradients thus leading to more imbalances in the pressure distribution. This implicitly induces more pressure gradient and the associated lift force thereby resulting significant augmentation in $\tilde{W}$. Besides, the modulation in the surface potential $(\gamma)$ strongly affects the flow velocity in the EDL region. Higher $\gamma$ means more is the strength of the electrokinetic force which generates higher pressure gradient and $\tilde{W}$ as a consequence.

For a fixed *p*H, increasing $n_0$ alters the surface charge density $(\Lambda)$ which follows a linear variation with Debye screening length $(\lambda_D)$, as clear from equation (S9) of the Supplementary Material. Meanwhile, $\lambda_D$ also depends on $n_0$, where it is defined as $\lambda_D^{-1} = \sqrt{\Omega e^2 n_0/\varepsilon}$. Thus, increasing $n_0$ gives rise to thinner $\lambda_D$ resulting in an increment in $\Lambda$ and $\zeta$. This enhancement in $\zeta$ is shown in the inset of figure 2. (a) where an increment up to ~ 4 times can be seen when $n_0$ is changed from 0.1 mM to 10 mM. This influences strongly the flow field close to the wall (the characteristic velocity scale $u_{HS}$ is directly proportional to $\zeta$, i.e. $u_{HS} \sim \varepsilon E \zeta / \mu$). Interestingly, this change is also reflected in degree of compliance factor $(\beta)$. Thus, increasing $n_0$ leads to significant enhancement in $\beta$. Physically, it strengthens the tendency of the plates to maintain the original separation between them which in turn induces more imbalances in the pressure distribution. Consequently, the generation of stronger *v*-component of flow field augments the lift force thereby resulting in an enhancement of $\tilde{W}/\tilde{W}_{ref}$ up to ~ 8.6 times, as can be seen from figure 2. (a) at higher $n_0$ (i.e. at $n_0$ = 10 mM). This figure also includes the effect of individual perturbations $\delta$ and $\gamma$. $\delta = 0.3$, $\gamma = 0$ represents modulated $l_s$ with axially invariant $\zeta$ where the strength of electrokinetic force gradually decreases as we increase $n_0$. Hence, any imbalance in pressure distribution is solely dictated by the modulated $l_s$ and gives rise to an increment in $\tilde{W}/\tilde{W}_{ref}$. However, in case of $\delta = 0$, $\gamma = 0.3$, the flow is governed by modulated $\zeta$ with constant $l_s$. With increasing $n_0$, the reduction in $\lambda_D$ makes the effect of $l_s$ to become inconsequential which, in the thin EDL limit, is often described by the Smoluchowski slip velocity. Intuitively, the enhancement in $\tilde{W}/\tilde{W}_{ref}$ gets dampened, i.e., an increment of ~ 2.8 times for $\delta = 0$, $\gamma = 0.3$, as opposed to ~ 16.6 times increment for $\delta = 0.3$, $\gamma = 0$. Mathematically, this can be explained using the definition of the function $\bar{Q}$ (this is involved in the respective velocity distribution) defined as $\bar{Q} = \sqrt{\bar{\kappa}^2 + \bar{q}^2}$. Thus, the reduction in $\lambda_D$ with increasing $n_0$ makes the effect of patterning frequency $(\bar{q})$ insignificant (i.e., $\bar{\kappa}^2 \gg \bar{q}^2$). The presence of both $\delta$ and $\gamma$ brings the interplay between the electrokinetic and wettability effects and the net enhancement of $\tilde{W}/\tilde{W}_{ref}$ depends on their relative strengths. Here, an increment of ~ 8.6 times is observed which lies in between their individual effects (~ 16.6 and ~ 2.8 times respectively).

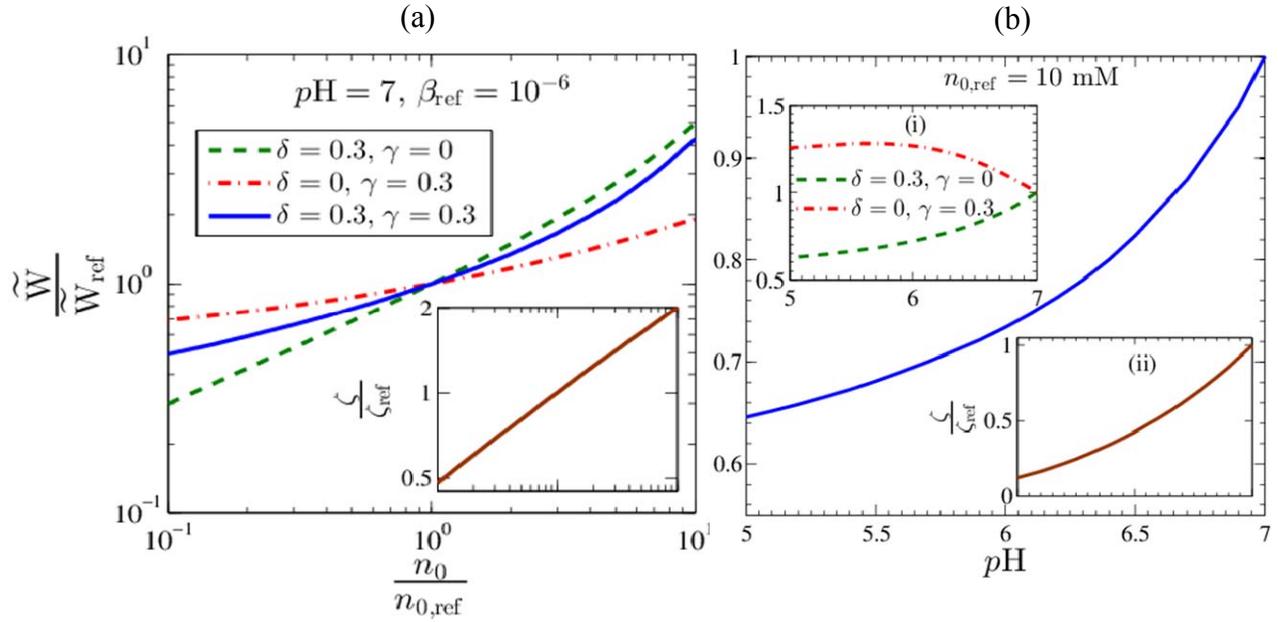

FIGURE 2. (a). The variation of $\tilde{W}/\tilde{W}_{ref}$ with ionic concentration $(n_0)$. Inset shows the variation of $\zeta$ with increasing $n_0$. $\tilde{W}_{ref}$ or $\zeta_{ref}$ corresponds to $n_0 = 1\,\text{mM}$. Fig. 2. (b). The reduction in $\tilde{W}/\tilde{W}_{ref}$ with $p$H. Inset (i) shows the effect of individual perturbations while inset (ii) corresponds to the variation of $\zeta$ with increasing $p$H. $\tilde{W}_{ref}$ or $\zeta_{ref}$ corresponds to $p\text{H} = 7$.

Now, for a fixed $n_0$ $(n_0 = 10\,\text{mM})$, decreasing $p$H from 7, more deprotonation reaction at the bulk results in lowering $\zeta$. This reduction is reflected in the inset (ii) of figure 2 (b) where a decrease of ~ 8.1 times has been observed as $p$H is changed from 7 to 5. Besides, the effects of individual perturbations on $\tilde{W}/\tilde{W}_{ref}$ are shown in inset (i) where two strikingly different trends can be noticed. For $\delta = 0.3$, $\gamma = 0$, the absence of axially varying component of $\zeta$ strengthens the effect of $l_s$ marked by significant alteration in the wall adjacent velocity gradient. The reduction of $\zeta$ with decreasing $p$H lowers the flow velocity and degree of compliance. This affects the pressure distribution and induces lower pressure gradient and $\tilde{W}/\tilde{W}_{ref}$ decreases up to ~ 1.6 times as compared to the case of $p\text{H} = 7$. However, for modulated $\zeta$ with constant $l_s$ (i.e. $\delta = 0$, $\gamma = 0.3$), two opposing dependences on $p$H is observed. This can be explained by following a close observation on the relative change in $\zeta$ with $p$H. In the regime of

$5.7 \leq p\text{H} \leq 7$, the reduction of $\zeta$ occurs much faster by following an exponential decaying behavior while within $5 \leq p\text{H} \leq 5.7$, it decreases slowly obeying a linear relationship. Since $\zeta$ is modulated, any change in it is strongly manifested in the flow field and the associated compliant behavior thereby enhancing $\tilde{W}/\tilde{W}_{ref}$ up to $p\text{H} \sim 5.7$, while beyond this critical $p\text{H}$, sluggish response in $\zeta$ slightly lowers $\tilde{W}/\tilde{W}_{ref}$. Nevertheless, in presence of both $\delta$ and $\gamma$ (i.e. $\delta = 0.3$, $\gamma = 0.3$), the effect of $l_s$ becomes dominant over $\zeta$ and a similar reduction of $\sim 1.55$ times is found. This is attributed to the shorter $\lambda_D$ and typically occurs at higher $n_0$ (reference $n_0$ is 10 mM). However, serial dilution of electrolyte can bring out a scenario when this two perturbations may become comparable to each other thereby resulting in the lower reduction of $\tilde{W}/\tilde{W}_{ref}$.

The influence of the phase-shift angle $(\theta)$ (or $\alpha$) on $\tilde{W}/\tilde{W}_{ref}$ is illustrated in figure 3 where $\tilde{W}_{ref}$ is the reference load bearing capacity at $\theta = 0°$ which means that the axially varying and invariant components are in the same phase. Here, it is important to highlight that, this kind of variation in $l_s$ is extremely important from design point of view where one can actually design the wettability gradient to control the degree of roughness of the surface. Using $\theta$ one can control the wall-adjacent flow field by imposing relative hydrophilic behavior in some region and hydrophobicity in some region thereby creating two distinct regions of lower and higher velocity gradients. This effect, when coupled with the compliant nature of surface, gives rise to an amplified imbalance in the pressure distribution. Such imbalance has an immediate effect in creating a disturbance in near-the-wall velocity distribution. More the degree of perturbation, more is the difficulty to propagate this disturbance into the outer layer through viscous interactions and hence leads to tremendous augmentation in the lift force and $\tilde{W}/\tilde{W}_{ref}$. For example, in case of axially modulated $l_s$ with constant $\zeta$, giant augmentation of $\tilde{W}/\tilde{W}_{ref}$ up to $\sim$ 26.5 times can be observed for $\tilde{\theta} = \pi/2$; i.e., when the axially varying and invariant components of $l_s$ are in $90°$ phase difference (clearly seen from figure 3. (i)). However, in presence of both $\delta$ and $\gamma$, the factor $\theta$ may bring out non-homogeneous consequences in the flow field. Depending on $\theta$, the electrokinetic and slip effects may act in tandem or counteract with each other. As $\theta$ is increased, the rapid growth of $\tilde{W}/\tilde{W}_{ref}$ gets attenuated a bit resulting in maximum increment of

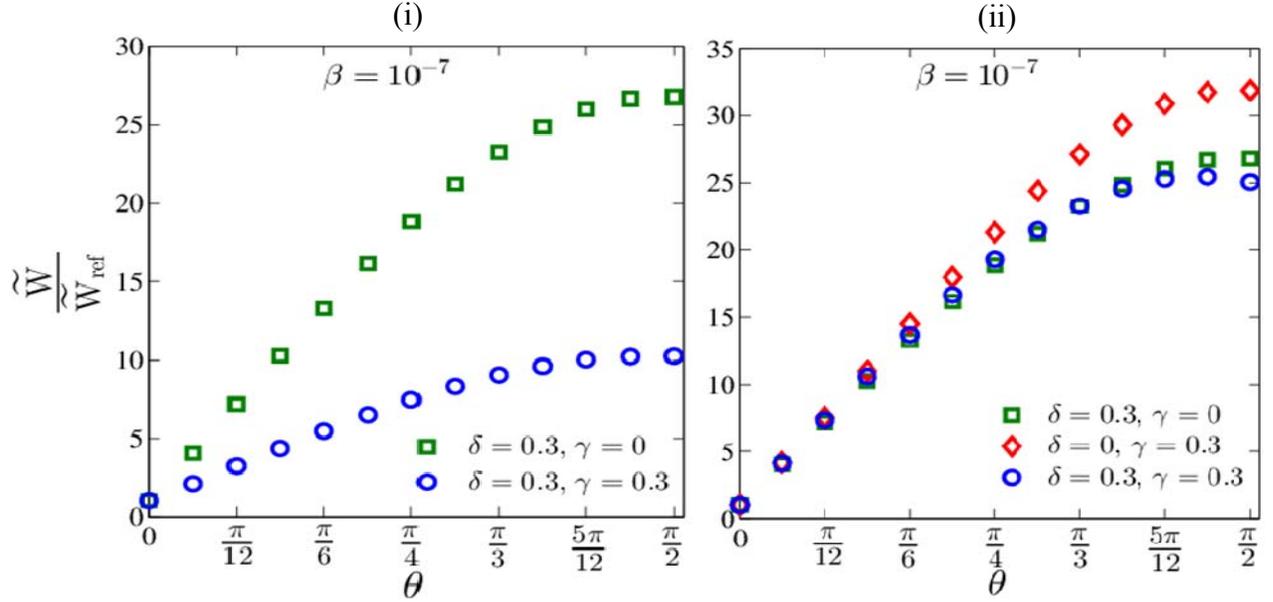

FIGURE 3. The enhancement in $\tilde{W}/\tilde{W}_{ref}$ with varying phase shift angle $\theta$ (or $\alpha$). (i) represents the case when the axially varying components of two modulations are in a phase difference while (ii) corresponds to the scenario at the same phase. Here $\theta$ and $\alpha$ are the phase-shift angles for modulated slip length and surface potentials respectively.

~ 10.25 times as attributable to the competition between two modulations, clearly visible in figure 3. (i). Additionally, employing a phase shift component in $\zeta$ drastically changes the amplitude of the enhancement in $\tilde{W}/\tilde{W}_{ref}$. Now, the axially varying component of $\zeta$ consists of two contributions, patterning frequency $(\bar{q})$ and angular contribution $(\theta)$. The relative strength of these two factors remain comparable only for smaller $\theta$, (up to $\theta = 30°$) while the periodic part becomes predominant at higher $\theta$, clearly manifested in figure 3 (ii) where modulated $\zeta$ with phase shift component results in ~ 31.9 times increase in $\tilde{W}/\tilde{W}_{ref}$. This figure also shows that the combinatorial effect of two perturbations is not simply linear superimposition of two effects, as the variation of $\tilde{W}/\tilde{W}_{ref}$ passes through a maximum around $\theta = 75°$ and then starts decaying. Also, the increment closely follows the variation with $\delta$ where any deviation can be noticed beyond $\theta = 75°$. For better understanding and completeness of this analysis, the relative deformations of the channel height in aforesaid cases are shown later.

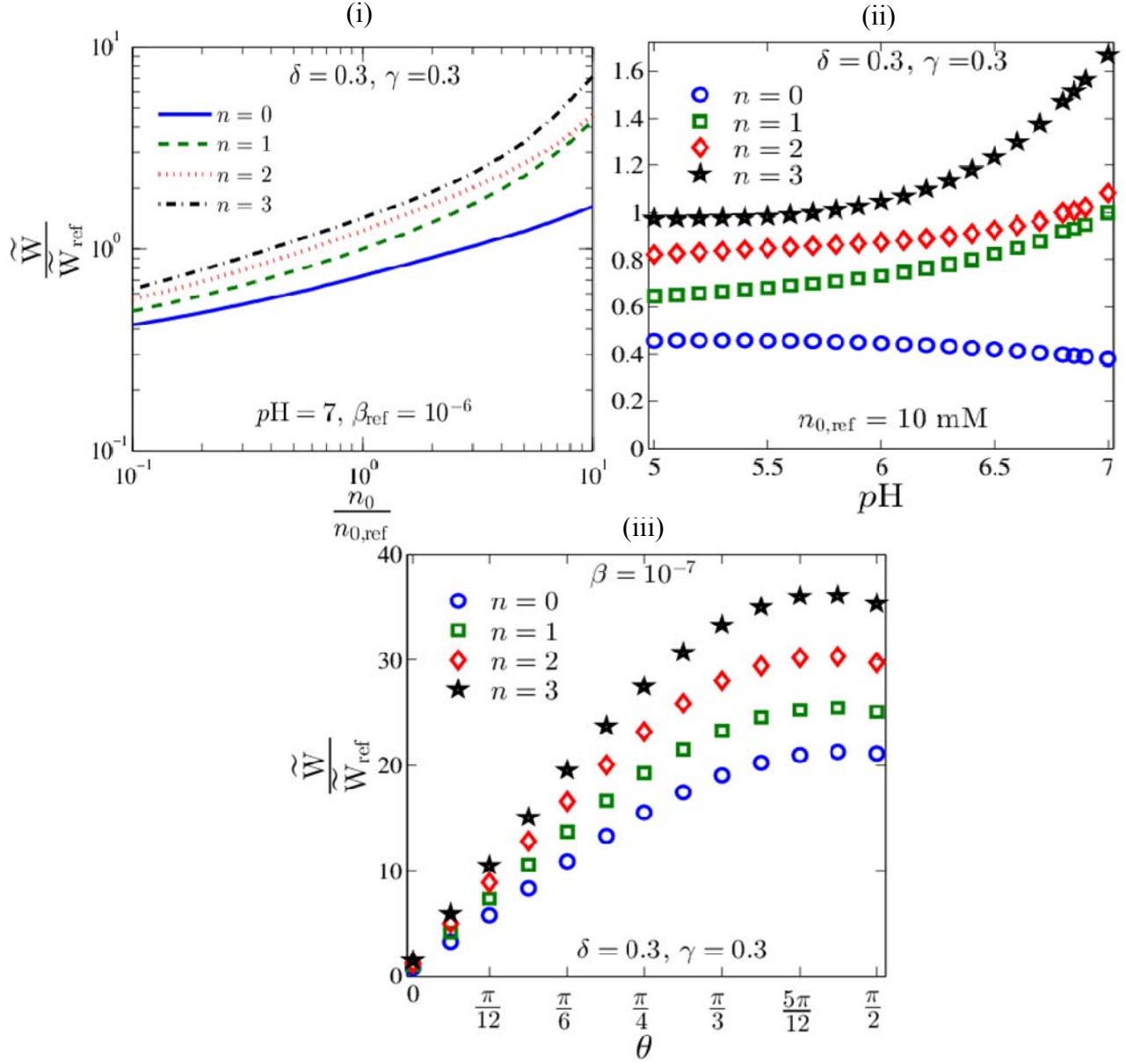

FIGURE 4. The variation in $\tilde{W}/\tilde{W}_{ref}$ for different values of $n$. (i) Varying $n_0$, (ii) varying $p$H and (iii) varying phase shift angle $\theta$.

The imposition of asymmetry in slip length $(l_s)$ is demonstrated in figure 4 where the variation of $\tilde{W}/\tilde{W}_{ref}$ is presented for the three previously mentioned conditions. As $n$ is increased, it introduces a relative increment in hydrophobicity of the surface. Higher the value of $n$, more is the perturbation in the wall-adjacent velocity gradient. So, more time is required to diffuse this disturbance into the outer layers thus resulting in significant increase in $\tilde{W}/\tilde{W}_{ref}$ for

$n = 3$. Similarly, decreasing $n$ creates a relatively hydrophilic region of reduced near-wall velocity thus affecting $\tilde{W}/\tilde{W}_{ref}$. One interesting thing to note that, decreasing $n$ value reverses the variation of $\tilde{W}/\tilde{W}_{ref}$ in case of $p$H alteration where the change in velocity gradient due to $\delta$ gets attenuated thus making surface charge distribution to govern the flow physics. Additionally, in presence of phase shift angle $(\theta)$ in the axially varying component, the amplitude of the change in flow field gets augmented with increasing $n$ and dampened with decreasing $n$, thus resulting ~ 35.3 times increase for $n = 3$ and ~ 21.1 times for $n = 0$.

For completeness of the present study, the respective deformations of the aforementioned cases are shown in figure 5 where the relative change in the channel height is depicted. As discussed previously, the variation of $d/h$ is strongly influenced upon increasing $n_0$, thus leading to considerable deformation for $n_0 = 10$ mM. Also, changing $p$H from 7 to 5 results in the reduction of $d/h$, although the effect is relatively weaker as compared to the effect of $n_0$. Also, imposing asymmetry $(n)$ considerably alters the extent of deformation, more the value of $n$, more is the deformation, clearly visible from figure 5 (iii). However, the influence of the phase difference $(\theta)$ between two modulations is strongest among these four alterations and consequently leads to giant augmentation in the associated load bearing capacity.

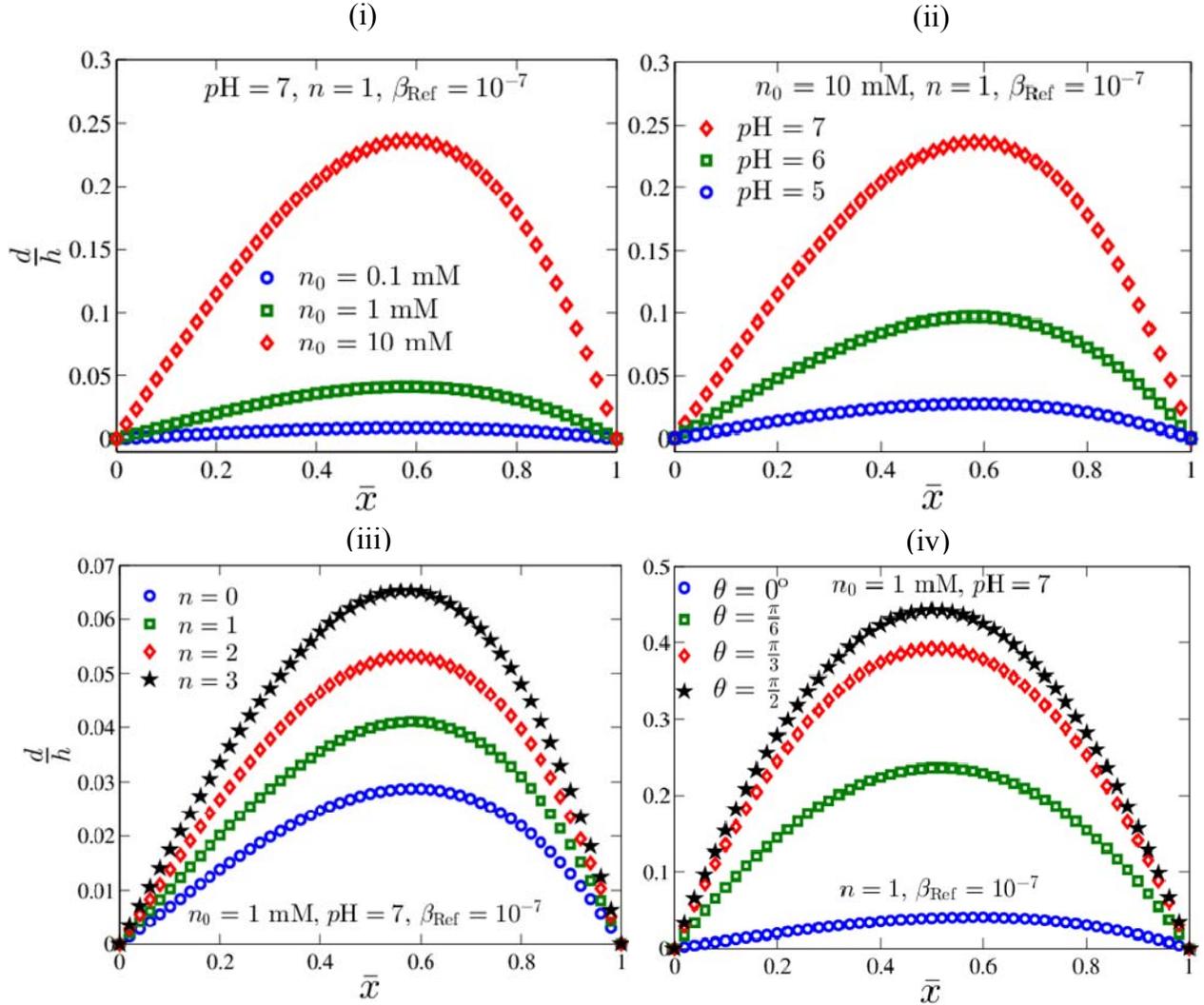

FIGURE 5. The axial variation of $d/h$ as a function of (i) varying $n_0$, (ii) varying $pH$, (iii) varying $n$ and (iv) varying $\theta$ where $\theta$ is the phase difference angle between two modulations.

For proper acknowledgement and visualization of the flow physics, we have depicted the contour plots for varying phase shift angle $\theta$ (or $\alpha$) in figure 6. Here, we have chosen to present the results only for phase shift angle since the degree of deformation becomes strongly influenced with $\theta$ as compared to the effects of other parameters. Physically, the combined effect of wettability and surface charge patterning basically establishes axial variation in the flow field which in turn induces a transverse component, and thus, one can notice the formation of vortices within the flow structure. However, when this two modulations act together, depending on the parameters involved, they may assist or counteract each other. In absence of phase-shift angle in the axially modulated component $(\theta = 0°)$, streamlines are slightly wavy in nature with where very small deformation is observed which mainly occurs from slip modulations (channel wall shows a little undulation in the axial direction for $\theta = 0°$). As we start increasing the value of $\theta$, one can observe recirculation patterns and vortex formation in the flow field. This is attributed to the fact that, although the periodic and other components are comparable for lower values of $\theta$, periodic component becomes predominant with increasing $\theta$ thus creating secondary flows within the domain. The induced transverse component of velocity gives rise to lift force to the compliant channel which undergoes deformation. At higher values of $\theta$, more is the undulation of the periodic component and hence, the strength of vortex formation is so pronounced that the channel experiences much higher deformation, clearly evident from the axial variation of the channel walls (for $\theta = \pi/3$ and $\theta = \pi/2$).

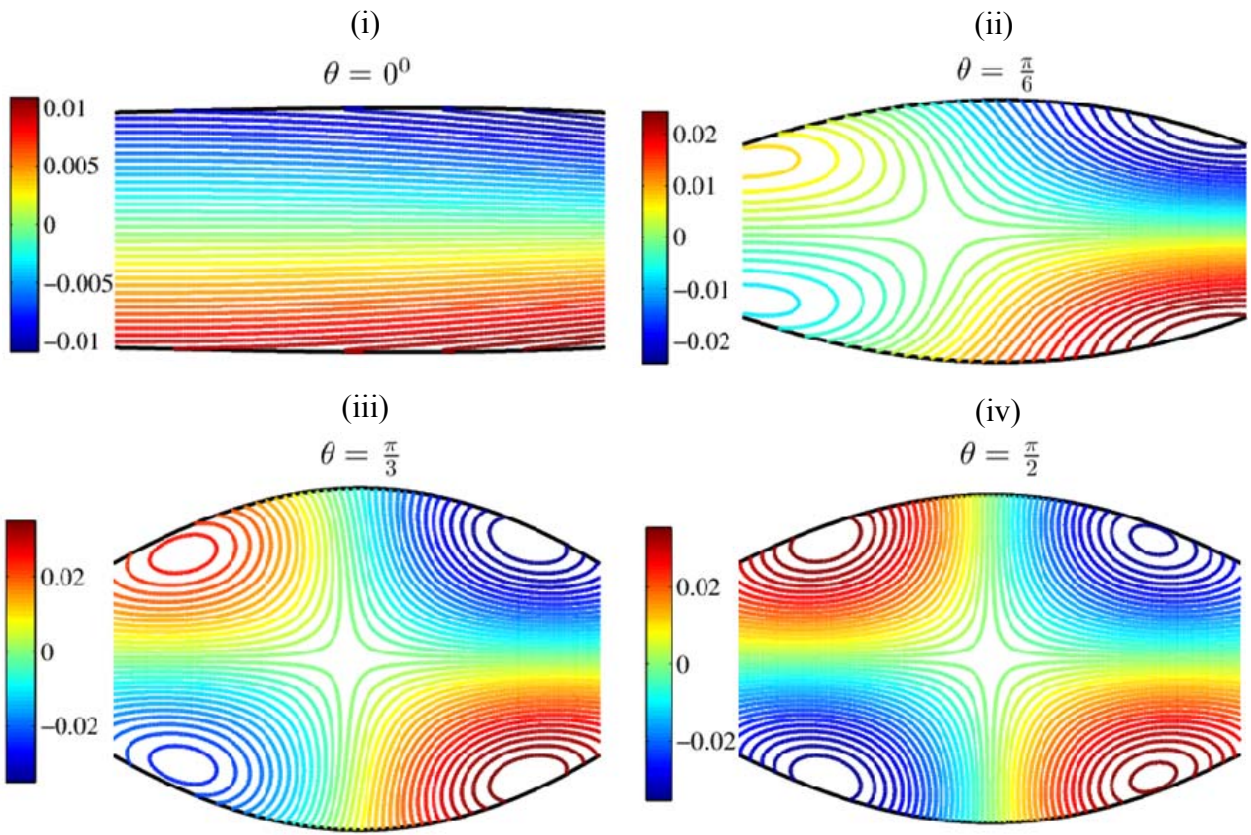

FIGURE 6. The dependence of the stream function for different phase shift angle $\theta$ (or $\alpha$) in the axial modulations. (i) $\theta = 0°$ (ii) $\theta = \pi/6$ (iii) $\theta = \pi/3$ and (iv) $\theta = \pi/2$. Results are presented for $n_0 = 10$ mM, $pH = 7$, $n = 1$ and $\beta = 10^{-7}$.

## 5. Experimental Perspective

In this section, we have presented practical relevance of the present analysis under real experimental conditions. For the fabrication of rectangular cross-section microchannel, one can employ the standard photolithography protocol which is followed by soft lithography and plasma bonding with the glass surface. Recently, PDMS-based microchannels are widely used owing to easier fabrication, functioning and compatibility. Another important factor is that the rheological property of these elastomers can be significantly altered upon doping with curing agents to the base, thereby altering the base to cross-link ratio. (Chakraborty *et al.* 2012; Del Giudice *et al.* 2016; Kang *et al.* 2014; Karan *et al.* 2018) This mixing with curing agent enhances the number of cross-links in the polymer chains thus exhibiting more viscoelastic behavior. (Raj *et al.* 2018, 2017) To carry out experiments, the measurement of the change in modulus of elasticity $(EI)$ can easily be done by performing the tensile test where the slope of the stress-strain curve characterizes the modulus of elasticity $(EI)$. For example, for PDMS (Sylgard 184, Dow Corning, Midland, MI, USA), changing the base to cross-link ratio from 10:1 to 30:1 results in reduction of $EI$ from 2.801 MPa to 0.157 MPa (i.e ~ 18 times). Following fabrication and property quantification, the measurement of microchannel deflection upon induced pressure gradient can be done using standard image processing methodologies like fluorescent imaging. For flow visualization, $\mu$-PIV technique can be implemented where a small concentration of fluorescent particles are seeded with the testing fluid.

To present the variation of $\tilde{W}/\tilde{W}_{\text{ref}}$ with varying $EI$, we have chosen the length scale $(L)$ in the longitudinal direction to be $\sim \text{O}(10^{-3})$ m while the half channel height $(H)$ and the slip length $(l_s)$ are chosen to be of the order of $\sim \text{O}(10)$ μm and $\sim \text{O}(100)$ nm, typically exists in microfluidic experiments. Considering the aforementioned alteration in $EI$ values, we have reported the results for $\bar{H} = H/L = 0.01$, $\bar{l}_s = l_s/L = 0.0001$, $\bar{q} = 5$ and $\theta = 0°$. Here, $\bar{q} = 5$ means that the patterning of surface charge and slip length are applied periodically 5 times in the axial direction while $\theta = 0°$ implies that modulations are in the same phase. The reference ionic concentration $(n_0)$ is chosen as 1 mM. As we start decreasing $n_0$, the electrokinetic effect on the flow field becomes more prominent and the reduction in $\zeta$ with $n_0$ becomes faster thus affecting the degree of compliance and the load bearing capacity $\left( \tilde{W}/\tilde{W}_{ref} \right)$. This ratio becomes half of its reference value upon changing $n_0$ from 1 mM to 0.1 mM, as evident from figure 7 (for $EI = 2.801$ MPa ). Now considering the aforesaid reduction in values of $EI$ $\left( EI = 0.157\,\text{MPa} \right)$ by changing base to cross-link ratio strongly influences the degree of compliance of the channel $(\beta)$. Lowering $EI$ enhances the parameter $\beta$ indicating more stiffness, thus resulting in the faster reduction of the load capacity than the case of $EI = 2.801$ MPa which yields ~ 1/4 of the ratio as compared to $\tilde{W}_{ref}$.

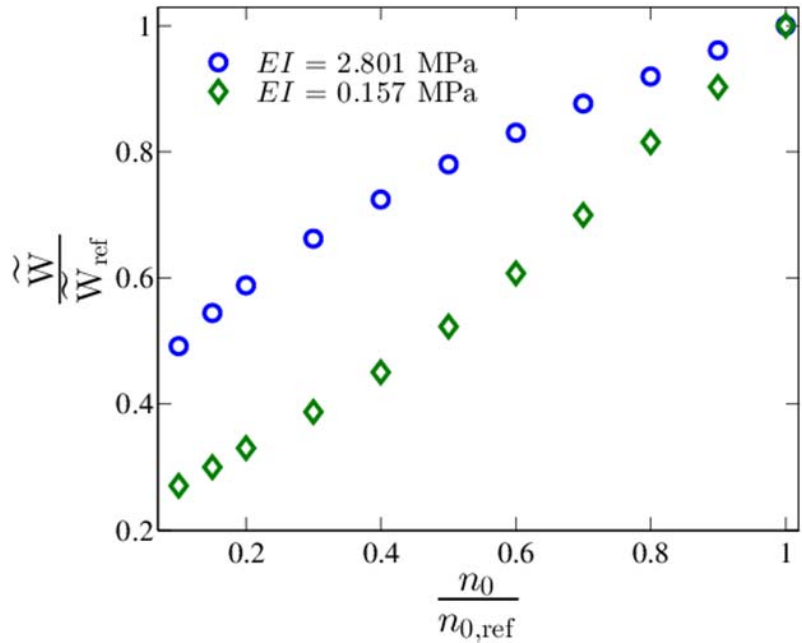

FIGURE 7. The dependence of $\tilde{W}/\tilde{W}_{ref}$ with $n_0$ for two different values of $EI$.

## 6. Conclusions

The central idea of the present analysis is to showcase the complex coupling between modulated hydrodynamic slip length and surface charge on the load bearing capacity of a deformable microchannel of parallel plate configuration. By taking into account the ionization of channel surface in contact with electrolyte solution, this study reveals that by judiciously tuning the pertinent parameters like ionic concentration $(n_0)$, pH, phase shift angle in axially varying component $(\theta)$ and enhancing asymmetry factor in slip modulation $(n)$, it is possible to achieve massive augmentation in the load bearing capacity of a compliant microfluidic channel. With rapid development in miniaturization technologies in recent years, the modulations of slip length and surface charge patterning considered here can easily be implemented by means of either some physical or chemical treatment (Bonaccurso *et al.* 2003; Dong *et al.* 2010; Jin *et al.* 2005; Kogoma *et al.* 1987; Sun *et al.* 2004). In this aspect, this analysis holds significant relevance in improved pragmatic design and performance of lubricated systems.

**Appendix**

**Section A: Solutions for $\mathrm{O}(\gamma)$ and higher degrees of perturbations:**

For $\mathrm{O}(\gamma)$, the potential distribution takes the following form

$$\psi_1 = \zeta \frac{\cosh(Qy)}{\cosh(QH)} \cos(qx+\alpha) \tag{A1}$$

where $Q$ is the modified patterning frequency $Q = \sqrt{q^2 + \kappa^2}$. Velocity profile takes the following form

$$\begin{aligned}
u_{12} = \frac{1}{2}\frac{d\bar{p}_{12}}{d\bar{x}}\left(\bar{y}^2 - \{\bar{H}_{12}(\bar{x})\}^2 - 2\bar{l}_s\bar{H}_{12}(\bar{x})\right) + \bar{l}_s\frac{\bar{\kappa}^2}{\bar{Q}}\tanh(\bar{Q}\bar{H})\cos(\bar{q}\bar{x}+\bar{\alpha}) \\
- \frac{\bar{\kappa}^2}{\bar{Q}^2}\cos(\bar{q}\bar{x}+\bar{\alpha})\left(\frac{\cosh(\bar{Q}\bar{y})}{\cosh(\bar{Q}\{\bar{H}_{12}(\bar{x})\})} - 1\right)
\end{aligned} \tag{A2}$$

where $\bar{H}_{12}(\bar{x}) = \bar{h} + \beta \bar{p}_{12}(\bar{x})$. Thus, the resulting differential equation for pressure distribution is given by

$$\begin{aligned}
b_1\frac{d^2\bar{p}_{12}}{d\bar{x}^2} + b_2\cos(\bar{q}\bar{x}+\bar{\alpha})\frac{d\bar{p}_{12}}{d\bar{x}} + b_3\sin(\bar{q}\bar{x}+\bar{\alpha})\left\{\frac{\tanh(\bar{Q}\bar{h})}{\bar{Q}} - (\bar{h}+\beta\bar{p}_{12})\right\} \\
+ b_4\sin(\bar{q}\bar{x}+\bar{\alpha})(\bar{h}+\beta\bar{p}_{12}) + b_5\left(\frac{d\bar{p}_{12}}{d\bar{x}}\right)^2(\bar{h}+\beta\bar{p}_{12}) = 0
\end{aligned} \tag{A3}$$

where $b_1 = a_1$, $b_2 = 2\beta b_6$, $b_3 = 2\bar{q}b_6$, $b_4 = -2\bar{l}_s\bar{q}b_7$, $b_5 = 2\bar{l}_s\beta$, $b_6 = \left(\frac{\bar{\kappa}}{\bar{Q}}\right)^2$ and $b_7 = \frac{\bar{\kappa}^2}{\bar{Q}}$.

For $\mathrm{O}(\delta^2)$, velocity profile:

$$u_{21} = \frac{1}{2}\frac{d\bar{p}_{21}}{d\bar{x}}\left(\bar{y}^2 - \{\bar{H}_{21}(\bar{x})\}^2 - 2\bar{l}_s\bar{H}_{21}(\bar{x})\right) - \bar{l}_s\bar{H}_{21}(\bar{x})\cos(\bar{q}\bar{x}+\bar{\theta})\frac{d\bar{p}_{11}}{d\bar{x}} \quad (A4)$$

where $\bar{H}_{21}(\bar{x}) = \bar{h} + \beta\bar{p}_{21}(\bar{x})$ and the corresponding pressure distribution is given in the following

$$c_1\frac{d^2\bar{p}_{21}}{d\bar{x}^2} + c_2\cos(\bar{q}\bar{x}+\bar{\theta})\frac{d^2\bar{p}_{11}}{d\bar{x}^2} + c_3\sin(\bar{q}\bar{x}+\bar{\theta})\frac{d\bar{p}_{11}}{d\bar{x}} + c_4\left(\frac{d\bar{p}_{21}}{d\bar{x}}\right)^2(\bar{h}+\beta\bar{p}_{21})$$
$$+c_5\cos(\bar{q}\bar{x}+\bar{\theta})(\bar{h}+\beta\bar{p}_{21})\frac{d\bar{p}_{11}}{d\bar{x}}\frac{d\bar{p}_{21}}{d\bar{x}} = 0 \quad (A5)$$

where $c_1 = b_1$, $c_2 = -2\bar{l}_s\bar{h}^2$, $c_3 = -\bar{q}c_2$ and $c_4 = c_5 = b_5$.

Similarly for $O(\gamma^2)$: velocity distribution:

$$u_{22} = \frac{1}{2}\frac{d\bar{p}_{22}}{d\bar{x}}\left(\bar{y}^2 - \{\bar{H}_{22}(\bar{x})\}^2 - 2\bar{l}_s\bar{H}_{22}(\bar{x})\right) \quad (A6)$$

with $\bar{H}_{22}(\bar{x}) = \bar{h} + \beta\bar{p}_{22}(\bar{x})$. Now, the pressure profile is given by

$$d_1\frac{d^2\bar{p}_{22}}{d\bar{x}^2} + d_2\left(\frac{d\bar{p}_{22}}{d\bar{x}}\right)^2(\bar{h}+\beta\bar{p}_{22}) = 0 \quad (A7)$$

with $d_1 = b_1$ and $d_2 = b_5$. It is noteworthy that the boundary condition $\zeta[1+\gamma\cos(qx+\alpha)]$ represents the modulation in the potential distribution up to $\sim O(\gamma)$ which means that for $O(\gamma^2)$, there is no effect in the charge distribution, pressure distribution remains unaffected, hence no deformation takes place.

For $O(\delta\gamma)$, flow field:

$$u_{23} = \frac{1}{2}\frac{d\bar{p}_{23}}{d\bar{x}}\left(\bar{y}^2 - \{\bar{H}_{23}(\bar{x})\}^2 - 2\bar{l}_s\bar{H}_{23}(\bar{x})\right) + \bar{l}_s\frac{\bar{\kappa}^2}{\bar{Q}}\tanh(\bar{Q}\bar{H})\cos(\bar{q}\bar{x}+\bar{\theta})\cos(\bar{q}\bar{x}+\bar{\alpha})$$
$$-\bar{l}_s\cos(\bar{q}\bar{x}+\bar{\theta})\bar{H}_{23}(\bar{x})\frac{d\bar{p}_{12}}{d\bar{x}} \quad (A8)$$

Pressure distribution:

$$e_1 \frac{d^2 \bar{p}_{23}}{d\bar{x}^2} + e_2 \frac{d^2 \bar{p}_{12}}{d\bar{x}^2} + e_3 \frac{d\bar{p}_{12}}{d\bar{x}} + e_4 \sin(\bar{q}\,\bar{x} + \bar{\theta})\cos(\bar{q}\,\bar{x} + \bar{\alpha})(\bar{h} + \beta\,\bar{p}_{23})$$

$$+ e_5 \cos(\bar{q}\,\bar{x} + \bar{\theta})\cos(\bar{q}\,\bar{x} + \bar{\alpha})\frac{d\bar{p}_{23}}{d\bar{x}} + e_6 \cos(\bar{q}\,\bar{x} + \bar{\theta})\sin(\bar{q}\,\bar{x} + \bar{\alpha})(\bar{h} + \beta\,\bar{p}_{23})$$

$$+ e_7 \left(\frac{d\bar{p}_{23}}{d\bar{x}}\right)^2 (\bar{h} + \beta\,\bar{p}_{23}) + e_8 \cos(\bar{q}\,\bar{x} + \bar{\theta})(\bar{h} + \beta\,\bar{p}_{23})\frac{d\bar{p}_{12}}{d\bar{x}}\frac{d\bar{p}_{23}}{d\bar{x}} \quad (A9)$$

$$+ e_9 \cos(\bar{q}\,\bar{x} + \bar{\theta})\cos(\bar{q}\,\bar{x} + \bar{\alpha})\frac{d\bar{p}_{23}}{d\bar{x}} = 0$$

where $e_1 = b_1$, $e_2 = c_2$, $e_3 = c_3$, $e_4 = b_4\,e_{10}$, $e_5 = b_5\,b_7\,e_{10}$, $e_6 = -2\,\bar{l}_s\,\bar{q}\,b_7\,e_{10}$, $e_7 = e_8 = b_5$, $e_9 = -b_5\,b_7\,e_{10}$ and $e_{10} = \tanh(\bar{Q}\,\bar{h})$. Hence, the total load carrying capacity of the compliant channel is given by

$$\tilde{w} = \tilde{w}_{11} + \tilde{w}_{12} + \tilde{w}_{21} + \tilde{w}_{23} = \delta \int_0^1 \bar{p}_{11}\,d\bar{x} + \gamma \int_0^1 \bar{p}_{12}\,d\bar{x} + \delta^2 \int_0^1 \bar{p}_{21}\,d\bar{x} + \delta\gamma \int_0^1 \bar{p}_{23}\,d\bar{x} \quad (A10)$$

**Section B: Stream function**

We use the definition of the stream function $\varphi$ and express it in the respective dimensionless forms as $\bar{u} = -\partial\bar{\varphi}/\partial\bar{y}$ and $\bar{v} = \partial\bar{\varphi}/\partial\bar{x}$ while $\bar{\varphi}$ is defined as $\bar{\varphi} = \varphi/L\,u_C$ where $u_C = u_{EOF}$ is the characteristic electroosmotic velocity scale. Now, to show the combined effect of two modulations $\delta$ and $\gamma$, we expand the stream function in a similar asymptotic series as described earlier:

$$\bar{\varphi} = \bar{\varphi}_0 + \delta\,\bar{\varphi}_{11} + \gamma\,\bar{\varphi}_{12} + \delta^2\,\bar{\varphi}_{21} + \delta\,\gamma\,\bar{\varphi}_{23} + \cdots \quad (B1)$$